\documentclass[a4paper,12pt]{article}
\newcommand{\be}{\begin{equation}}
\newcommand{\ee}{\end{equation}}
\newcommand{\ba}{\begin{eqnarray*}}
\newcommand{\ea}{\end{eqnarray*}}

\usepackage{graphicx,pstricks}
\usepackage{setspace}
\usepackage {amssymb,amsmath}
\usepackage{parskip}
\title{Decaying Holographic Dark Energy and Emergence of Friedmann Universe}
\author{Titus K Mathew \\ 
 Department of Physics, \\
Cochin University of Science and Technology, 
\\{\it  Kochi-22, India}.\\
\vspace{0.2in}
E-mail: tituskmathew@gmail.com, titus@cusat.ac.in.}
\date{}
\begin{document}

\maketitle

\begin{abstract}
  A universe started in almost de Sitter phase, with time varying holographic dark 
energy corresponding to a time varying cosmological term is considered. The time varying cosmological 
dark energy and the 
created matter are consistent with the Einstein's equation. The general conservation law for the decaying
dark energy and the created matter is stated. By assuming that the initial matter were created
is in relativistic form, we have analyzed the possibility of evolving the universe from de Sitter phase to Friedmann
universe.

\end{abstract}

Keywords: dark energy, Friedmann Universe, cosmology
\section{Introduction}

Recent Astrophysical data have shown that the present universe is undergoing an 
accelerated expansion \cite{Perl98}. This shows that the present universe is 
dominated by some kind  of very smooth form of energy with negative pressure 
has been called dark energy, which accounts for about 75$\%$ of the total energy density
of the present universe. Various models have been proposed to explain this phenomenon, 
for example there are models based on the dynamics of scalar or multi-scalar filed, 
called quintessence models \cite{rathra88}. Another dark energy candidate is the 
cosmological constant, which was initially introduced by Einstein. 
In the cosmological constant model, 
the dark energy density, $\rho_{\Lambda}$, remains constant throughout the entire 
history of the universe, while the matter density decreases during the expansion.
 The equation of the state for cosmological constant as dark energy is 
$ w = p/ \rho_{\Lambda} = -1$. While in Phantom models \cite{Huang05}, it is possible to 
have an equation of state with $w < -1.$

An alternative approach to the dark energy problem arises from the holographic 
principle. According to the principle of holography the number of degrees of 
freedom in a bounded system should be finite and has relations with area of 
its boundary.  By applying the principle to cosmology, one can obtain the 
upper bound of the entropy contained in the universe. For a system with 
size $L$ and UV cut-off $\Lambda$, without decaying into a black hole, it is 
required that the total energy in a region of size $L$ should
not exceed the mass of a black hole of the same size, thus $L^3 \rho_{\Lambda}
 \leq L M_P^2$ . The largest L
allowed is the one saturating this inequality, thus
\begin{equation} \label{eq:rho1}
 \rho_{\Lambda} = 3c^2 M_P^2 L^{-2}
\end{equation}
where $c$ is numerical constant having value close to one, we will take it as one in our analysis.
and $M_P$ is the reduced Planck Mass $M_P^{-2} = 8 \pi G.$   When we take 
the whole universe into account,the vacuum energy related to this holographic
 principle can be viewed as dark energy.  $L$ can be taken as the large scale 
of the universe, for example Hubble horizon, future event horizon or particle 
horizon which were discussed by many \cite{hsu04,Li04,Huang05,horava2000,Horavat04}.

In this paper we assume a decaying cosmological term. We also assume that the universe is started in 
de-Sitter phase. While in the de-Sitter phase the universe is completely dominated with the cosmological term.
As the universe expands the dynamical cosmological term decaying in to matter and the universe will
subsequently enter the Friedman phase. As it expands further, the universe enter a matter dominated phase with 
decelerated expansion. In section two we have shown that
the decaying cosmological dark energy and created matter are consistent with the Einstein's equation. In section 
3, we have obtained the Friedmann equations for the decaying dark energy, and analyzed the possibility of 
the evolution of the universe in to the Friedmann phase. We have also obtained the time evolution of the 
decaying dark energy and its equation of state. In section 4, we have presented a comprehensive discussion of
our analysis.

\section{Dynamical dark energy and horizon}

        In the presence of cosmological constant the Einstein's field equation is
\begin{equation} \label{eq:ein1}
G^{\mu \nu} = R^{\mu \nu} - \frac{1}{2} R g^{\mu \nu} = {8\pi G \over c^4} T^{\mu \nu}_{total}
\end{equation}
where $G^{\mu \nu}$ is the Einstein tensor, $R^{\mu \nu}$ is Ricci tensor, $R$ is the Ricci scalar (except in this
equation, we will refer $R$ as the scale factor of the expanding universe)and
$T^{\mu \nu}_{total}$ is the total energy momentum tensor comprising matter and cosmological term, and is
\begin{equation}
 T^{\mu \nu}_{total} = T^{\mu \nu} + \rho_{\Lambda} g^{\mu \nu}
\end{equation}
in which $T^{\mu \nu}$ is the energy momentum tensor due to matter in perfect fluid form and $\rho_{\Lambda}$
is the density due to cosmological term, given as,
\begin{equation}
 \rho_{\Lambda} = {c^4 \Lambda \over 8 \pi G}
\end{equation}
with $\Lambda$ as the so called ``cosmological constant''. 

      Einstein's equation satisfies the covarient conservation condition, 
\begin{equation}
 \nabla_{\mu} G^{\mu \nu} = 0
\end{equation}
In the conventional case this implies that, $\nabla_{\mu} T^{\mu \nu} = 0.$ As such this 
condition doesn't give any time conserved charge. If the matter is being created form an independent
source, say form the cosmological term, the conservation law will then take the general form
\begin{equation}
 \nabla_{\mu} \left( T^{\mu \nu} + \rho_{\Lambda} g^{\mu \nu} \right) = 0
\end{equation}
This conservation law implies that the energy and momentum of matter alone is not conserved, bur 
energy and momentum of matter and cosmological term or dark energy are together be conserved. This 
general conservation law allows the exchange of energy and momentum between matter and dark energy and acting
as a controlling condition for this exchange. The existing theories predicts a very large value for the 
cosmological term \cite{weinb89, Linde90} in the early stage of the universe, but the present 
observations points towards a very low value for the cosmological term for the late universe. In this
light it is inevitable to consider that, there must be a transference of energy form the dark energy or 
cosmological term sector to the matter sector.

Let us assume that, the term $\Lambda$ correspondingly $\rho_{\Lambda}$ is a function of time, since a space
dependent $\Lambda$ will lead to an anisotropic universe. The covarient conservation law will then give the 
equations,
\begin{equation}
 \nabla_{\mu} T^{\mu i} = 0
\end{equation}
and 
\begin{equation} \label{eq:con1}
 \nabla_{\mu} T^{\mu 0} = -{c^3 \over 8\pi G} {d\Lambda \over dt}
\end{equation}
where $i=1,2,3$ for the spatial part and $i=0$ for the time part.

In reference \cite{abp2005} authors have considered the energy transference between
decaying cosmological term and  matter. It is important to realise that the covarient conservation law 
given above is drastically different from that appearing in in some quintessence model
\cite{ratra1988, friman1992, sahni2000}, where energy-momentum tensor of the scalar field 
that replaces the cosmological term is itself covariently conserved, but no matter creation. 
In the present paper we have considered that the cosmological term decaying into matter
which is consistent with the Friedmann model of the universe.

    The energy density $\rho_{\Lambda}$ corresponds to the time varying cosmological 
term is taken as the holographic dark energy as defined in equation (\ref{eq:rho1}).
 A simple holographic 
dark energy model is by taking $L= H^{-1}$, where $H$ is the Hubble's constant is considered by Hsu et al 
\cite{hsu04} and they have shown that the Friedmann model with $\rho_{\Lambda} = 3 c^2 M^2_p H^2 $ 
makes the dark energy behave like ordinary matter rather than a negative pressure fluid, 
and prohibits accelerating expansion of the universe. We have adopt an equation for 
holographic dark energy energy, where the future event horizon $(R_h )$ is used instead
of the Hubble horizon as the IR cut-off L, which was shown to lead an accelerating universe
by Li \cite{Lietal2004}. Thus the time varying cosmological energy density is 
\begin{equation} \label{eq:rho2}
 \rho_{\Lambda} = 3 c^2 M^2_p R_h^{-2} 
\end{equation}
where $c$ is a constant have values $O(1)$ and the event horizon $R_h(t)$, 
a function of cosmological time, is given by
\begin{equation} \label{eq:Rh1}
 R_h (t) = R(t) \int^{\infty}_t {dR(t^{'}) \over H(t^{'}) R(t^{'})^2 }
\end{equation}
where $R(t)$ is the expansion factor and $H(t)$ is the Hubble constant. 

\section{Cosmic evolution of dark energy and Friedmann Universe}

    Let us consider an empty universe in de Sitter phase, with very large \cite{mashhoon2004} decaying 
cosmological term, corresponds to dark energy density as given by equation (\ref{eq:rho12}). If the matter 
and energy are created 
from the decaying dark energy term are homogeneous and isotropic, then the geometry 
of the universe can be that of Friedmann-Robertson-Walker form,
\begin{equation} \label{eq:frw}
 ds^2 = c^2 dt^2 - R^2 \left[ {dr^2 \over 1-kr^2} + r^2 \left(d\theta^2 + 
\sin^2\theta \, d\phi^2 \right) \right]
\end{equation}
where $k$ is the curvature parameter $R$ is the scale factor of expansion, $t$ is the cosmological time 
and $(r,\theta, \phi) $ are the co-moving coordinates.
By taking the energy momentum tensor of matter as 
\begin{equation} \label{eq:tmu1}
 T^{\mu}_{\nu} = (\rho + p) u^{\mu} u_{\nu} - p \delta^{\mu}_{\nu}
\end{equation}
where $\rho $ is the sum of energy densities of the created 
components of due to the decay of cosmological term and $p $ 
is the pressure of the matter components. Under these conditions, the covarient conservation law 
(\ref{eq:con1})
leads to (here we consider only one component of matter)
\begin{equation} \label{eq:con2}
{d \rho_m \over dt} + 3H \left ( \rho_m + 
p_m \right) = - {d\rho_{\Lambda} \over dt}
\end{equation}
where $H={dR \over dt} / R$ the Hubble parameter, $\rho_m$ is the density of the created matter and $p_m$ is its
pressure. This equation 
obtained form the general conservation law is found to followed from the combinations the standard Friedmann
equations,
\begin{equation} \label{eq:fried1}
 \left( {dR \over dt} \right)^2 = {8 \pi G \over 3c^2} \left( \rho_m + \rho_{\Lambda} \right) R^2 - k R^2
\end{equation}
and
\begin{equation} \label{eq:fried2}
 {d^2 R \over dt^2} = {8 \pi G \over 3c^2} \left( \rho_{\Lambda} - \frac{1}{2} 
\left(\rho_m + 3 p_m \right) \right) R
\end{equation}
provided the cosmological term $\rho_{\Lambda}$ is time dependent.
If one assumes  ordinary pressureless matter as
\begin{equation} \label{eq:Rh3}
  \rho_m = \rho_{m0} R^{-3}
\end{equation}
where $\rho_{m0}$ is the present density of matter.
then equation (\ref{eq:con2}) will lead to the result that, the cosmological term will be 
independent of time. On the other hand this shows that the time dependent cosmological
term does not decay in to pressureless matter.

Let us assume that the cosmological term can possibly decay into some form of matter
with equation of state $p_m = \omega_m \rho_m, $ where the parameter $\omega_m$ is assumed to be in the range 
$0 \leq \omega_m \leq 1,$ the exact value is depends on particular matter component. In this paper we are 
considering only one component of matter. The covarient conservation law (\ref{eq:con1}) can now be written for the 
possibility of cosmological 
term decaying into matter  as
\begin{equation} \label{eq:con3}
 {d \rho_m \over dt} + 3H \left(1 + \omega_m \right) \rho_m = -{d\rho_{\Lambda} \over dt}
\end{equation}
Since density behaviour of ordinary pressureless matter does not work for a varying cosmological dark energy,
we will assume the form for $\rho_m$ which is slightly different from its canonical form,
as \cite{horvat2007, horvat2006}
\begin{equation} \label{eq:rhoom1}
\rho_m = \rho_{m0} R^{-3 + \delta}
\end{equation}
where $\rho_{m0}$ is the present value of $\rho_m$ and $\delta$ is a parameter which is effectively depends upon the
state of the universe. From 
equation (\ref{eq:rhoom1}) and (\ref{eq:rho12}), equation (\ref{eq:con3}) become,
\begin{equation} \label{eq:rhorelatns}
 \rho_{\Lambda} \left ( {HR_h -1 \over H R_h} \right) = \left( {3 \omega_m + \delta \over 2} \right ) \rho_m
\end{equation}
where we have assumed a vanishing integration constant and also with the condition that 
lim t$\rightarrow \infty$, $R(t) \rightarrow \infty $ 
and equation for time rate of horizon can be cast into the diﬀerential form 
\begin{equation} \label{eq:Rhderi}
 {d R_h \over dt} = H R_h - 1.
\end{equation} The equation (\ref{eq:rhorelatns}) suggest that 
depending on the parameters $\delta$ and $\omega_m$ the energy densities $\rho_m$ and $\rho_{\Lambda}$
may eventually be of the same order, as suggested by the present observations \cite{smcarrol}.
This relation rather suggest the relation between dark energy and matter, when 
$\delta = 3$, the case corresponds to a constant
matter density at which there exist a equilibrium between matter creation and universe expansion

In general event horizon is not existing for Friedmann universe. But for de Sitter universe there exists event 
horizon, which satisfies the relation $R_h \sim H^{-1}.$ Consequently for de Sitter universe, $HR_h \sim 1$
which implies that in de Sitter phase the energy density is almost completely dominated by the cosmological term
or dark energy. For Friedmann universe we will hence take $HR_h$ as very large  
 In general we will take the value of $HR_h$ is equal to one or large.

  \subsection {Friedmann Universe}

Friedmann universe is a homogeneous and istropic universe, satisfying the conditions (\ref{eq:fried1}) and 
(\ref{eq:fried2}). 
With the relation between decaying cosmological term and created matter (\ref{eq:rhorelatns}), the second Firedmann
equation become,
\begin{equation}  \label{eq:Rrate1}
 {d^2R \over dt^2} = {( \left(1+3 \omega_m \right) \beta^2 \over 2} 
\left[ {3 \omega_m + \delta 
\over 1+3 \omega_m} \left({HR_h \over HR_h-1} \right) - 1 \right] R^{\delta -2}
\end{equation}
where $\beta^2 = {8\pi G \rho_{mo}/3c^2}$ . For de Sitter phase the acceleration is very large, for which 
$HR_h \sim 1.$ As it enters the Friedmann phase by the decay of cosmological or the dark energy, the acceleration 
can be negative or positive, depends on the value of the term in the parenthesis of the right hand side 
of the above equation. The condition for acceleration is,
\begin{equation}\label{eq:cond1}
 {3\omega_m + \delta \over 3 \omega_m + 1} > 1- {1 \over HR_h}
\end{equation}
The factor $\displaystyle {1- {1 \over HR_h}}$ is in the range $\displaystyle 0 \leq {1- {1 \over HR_h}} \leq 1.$
The extreme limits are corresponds to de Sitter phase and matter dominated universes respectively. For the 
transition period from de Sitter phase to Friedmann phase, $HR_h$ is near to one, and assuming that the 
created matter have the equation state $\omega_m = \frac{1}{3}$, where the created matter is in relativistic 
form then
\begin{equation}
 \delta > 2\alpha - 1
\end{equation}
where $\displaystyle \alpha=1-{1 \over HR_h}$ having value less than one. This implies that during the period 
of decay of the cosmological dark energy term
the density of created matter is diluted slowly as the universe expand, than the decreasing of the 
density of non-relativistic matter in the Friedmann universe. This shows that even in the friedmann universe 
it is possible to have an initial accelerating phase, where the cosmological dark energy is start its 
decay into matter and is still dominating over matter. As
universe proceeds, the created matter will subsequently dominate and hence the universe will come to a matter dominated
phase, at which the universe is expanding with deceleration.

For decelerating expansion, where matter is dominating over the cosmological term, the condition in the limit where
$HR_h$ is very large is,
\begin{equation}
 \delta < 1
\end{equation}
This condition is true irrespective of whether the created matter is relativistic or non-relativistic. However as the 
universe enter the decelerating phase, the matter will become non-relativistic, satisfying the extreme condition that
$\delta \rightarrow 0$ so that $\rho_m \sim R^{-3}$

\subsection { Flat Universe}
For flat universe, where the curvature parameter $k=0,$ the Friedmann equation (\ref{eq:frw1}) become
\begin{equation}
 \left({dR \over dt} \right)^2 = \beta^2 \left( {3\omega_m + \delta \over 2} {HR_h \over HR_h - 1} + 1 \right) R^{\delta -1}
\end{equation}
On integration and avoiding the integration constant, the solution would be,
\begin{equation}
 R \sim t^{2\over 3 - \delta}
\end{equation}
By considering the relation for matter creating out of decaying dark energy, i.e. $\rho_m = \rho_{m0} R^{-3 + \delta},
$
then the relation between dark energy density and matter will have beheaviour
\begin{equation}
 \rho_m \sim \rho_{\Lambda} \sim t^{-2}
\end{equation}
This is the time dependence of $ \rho_{\Lambda}$ for any value of $\omega_m$ and $\delta.$ 
This time dependence shows that $\rho_{\Lambda}$ diverge at the initial time, which implies the existence of 
initial singularity.

\subsection {An equation of state for the decaying dark energy}

An equation of state for the time decaying cosmological term can be written as \cite{fang2008}
\begin{equation}
 \omega_{\Lambda}^{eff} = -1 - {1 \over 3} \left( {d ln \rho_{\Lambda} \over d \ln R} \right)
\end{equation}
With the equation for the relation between decaying dark energy and creating matter \cite{fang}, the equation of state become
\begin{equation}
 \omega_{\Lambda}^{eff} = -{\delta \over 3} - { 1 \over 3 \left( HR_h - 1 \right)^2}
\end{equation}
This shows that for large values of $HR_h$ the equation of sate become $\omega_{\Lambda} = -{\delta \over 3}$.
From the above analysis, it is seen that for an accelerating universe $\omega_{\Lambda}$ is less than $-{1 \over 3},$
but for a decelerating Friedmann universe it is greater than $-{1 \over 3},$ which is similar to the latest 
analysis by many.

\section{Discussion}
In the presence of a time varying cosmological term, assumed to be of holographic dark energy form, it is 
possible that the universe may starts with the de Sitter phase, exhibiting horizon and 
where the energy density is completely 
dominated by the dark energy. The decaying holographic dark energy cause the primordial inflation. If the horizon 
$R_h$ is assumed to be equal to the plank length at very early stage, then $\Lambda$ would have a value of the order
\begin{equation}
 \Lambda \sim 10^{66} \, cm^{-2}
\end{equation}
the corresponding dark energy density would be $\rho_{\Lambda} \sim 10^{112} \, erg cm^{-3}.$ This enormous dark energy
decay into matter as the universe is evolved to the Friedmann phase, and the dark energy reached the present value
\begin{equation}
 \rho_{\Lambda}^0 = 10^{-8} \, erg \, cm^{-3}
\end{equation}
The evolution of the de Sitter phase in to the Friedmann universe is in such a way that the total energy density
comprising the dark energy, created matter and the gravitational field together be conserved. In this paper we have 
considered the decay of the dark energy into matter. During the initial phase of decay, the universe might be in the 
accelerating phase, where the parameter $\delta$ characterizing the equation $\rho_m \sim R^{-3+\delta},$ is 
greater than one. This implies that the dilution in the density of created matter in slower compared to the 
non-relativistic matter. As the universe proceeds with expansion, the matter created will come to dominate, 
and the universe eventually go over to matter dominated phase, with the condition $\delta < 1.$ In the extreme limit
this condition may go the limit $\delta \rightarrow 0$, which emphasises that, the created matted will eventually 
become non-relativistic, with behaviour, $\rho_m \sim R^{-3}.$ In section 3.3, the equation of the state of 
the decaying holographic dark energy, shows that, in the early phase of universe too, 
$\omega_{\Lambda} < -{1 \over 3}$ as it's equation of state in late accelerating universe \cite{sami2008}.
To explain why 
$\rho_{\Lambda}$ and $\rho_m$ are of the same order today, it is essential to have  a specific time evolution for
dark energy. We argued that a dynamical dark energy, endowed with an appropriate time evolution can 
contain the possibility of the development of a Friedmann universe from a de Sitter universe. As the de Sitter 
phase evolved in to the Friedmann universe, the value of the dark energy is decreased gradually to a low value, 
which eventually lead to matter dominating phase with decelerating expansion. But on the other hand the recent
observations indicating that the present universe is in a accelerating expansion. This fact indicating the
possibility that at present the dark energy is increasing at the expense of decaying matter. This time increasing
 dark energy, in other words, implies that the universe may evolving in to stage where the whole energy density is coming
to dominate completely by the dark energy. The ultimate clarity regarding these, of course be given by the proper quantum effects,
which is still an open question.

\end{document}